\documentclass[sigconf]{acmart}

\AtBeginDocument{%
  \providecommand\BibTeX{{%
    \normalfont B\kern-0.5em{\scshape i\kern-0.25em b}\kern-0.8em\TeX}}}

\setcopyright{acmcopyright}
\copyrightyear{2024}
\acmYear{2024}
\setcopyright{rightsretained}
\acmConference[DIS '24]{Designing Interactive Systems Conference}{July 1--5, 2024}{IT University of Copenhagen, Denmark}
\acmBooktitle{Designing Interactive Systems Conference (DIS '24), July 1--5, 2024, IT University of Copenhagen, Denmark}\acmDOI{10.1145/
XXXXXXX.XXXXXXX}
\acmISBN{979-8-4007-XXXX-X/24/07}

\begin{document}

\title{Conversational Agents as Catalysts for Critical Thinking: Challenging Design Fixation in Group Design}





\author{Soohwan Lee$^{*}$}
\affiliation{%
  \institution{Department of Design, UNIST}
  \city{Ulsan}
  \country{Republic of Korea}}
\email{soohwanlee@unist.ac.kr}

\author{Seoyeong Hwang$^{*}$}
\affiliation{%
  \institution{Department of Design, UNIST}
  \city{Ulsan}
  \country{Republic of Korea}}
\email{hseoyeong@unist.ac.kr}

\author{Kyungho Lee}
\affiliation{%
  \institution{Department of Design, UNIST}
  \city{Ulsan}
  \country{Republic of Korea}}
\email{kyungho@unist.ac.kr}

\thanks{$^{*}$ Equally contributed to this work.}

\renewcommand{\shortauthors}{Lee et al.}

\begin{abstract}
This paper investigates the potential of LLM-based conversational agents (CAs) to enhance critical reflection and mitigate design fixation in group design work. By challenging AI-generated recommendations and prevailing group opinions, these agents address issues such as groupthink and promote a more dynamic and inclusive design process. Key design considerations include optimizing intervention timing, ensuring clarity in counterarguments, and balancing critical thinking with designers' satisfaction. CAs can also adapt to various roles, supporting individual and collective reflection. Our work aligns with the "Death of the Design Researcher?" workshop's goals, emphasizing the transformative potential of generative AI in reshaping design practices and promoting ethical considerations. By exploring innovative uses of generative AI in group design contexts, we aim to stimulate discussion and open new pathways for future research and development, ultimately contributing to practical tools and resources for design researchers.
\end{abstract}



\begin{teaserfigure}
  \includegraphics[width=\textwidth]{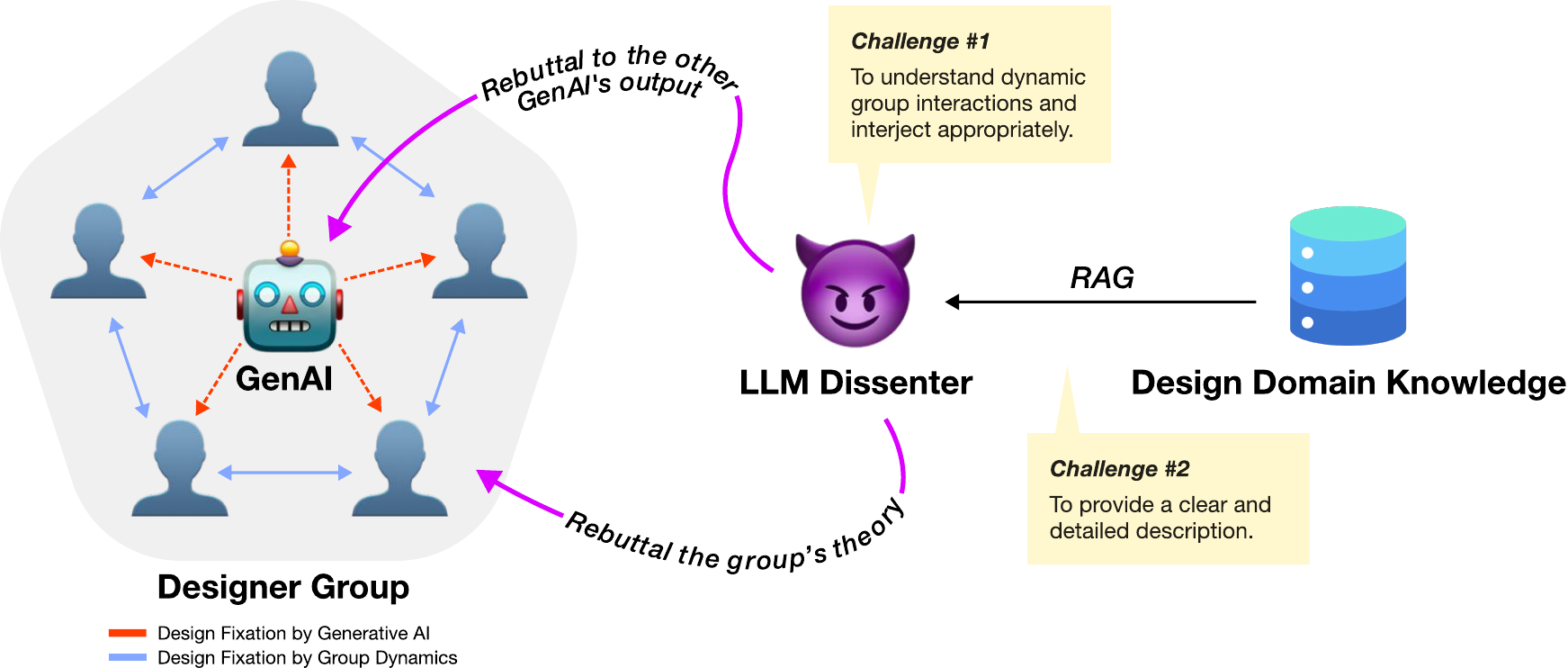}
  \caption{Illustration of an LLM-based Conversational Agent (CA) Acting as a Dissenter in Group Design. This work-in-progress shows how a CA can promote critical thinking by 1) challenging recommendations from other generative AIs and 2) opposing prevailing group opinions. Key challenges for CAs in group discussions include understanding dynamic interactions to decide optimal intervention times and providing clear, detailed counterarguments. We propose linking design domain knowledge with Retrieval-Augmented Generation (RAG) techniques to address these challenges.}
  \Description{}
  \label{fig:teaser}
\end{teaserfigure}


\maketitle

\section{Introduction}
The Candle Problem, devised by Gestalt psychologist Karl Duncker in 1945, is a classic cognitive experiment designed to assess creative problem-solving and demonstrate the concept of functional fixedness \cite{duncker_functional_2019}. It reveals how our preconceived notions about the function of objects can hinder our ability to find creative solutions. The solution requires a shift in perspective, where the box of thumbtacks is seen as a container and a potential platform for the candle. This experiment highlights the importance of overcoming functional fixedness by mentally restructuring the problem to see objects in new and unconventional ways. Such creative thinking is particularly crucial in the design process, especially during the early ideation and divergence phases when generating a wide array of ideas is essential \cite{alipour_review_2018}.

However, in the context of design, whether in practice or research, designers often struggle to engage in meaningful self-reflection or constructive criticism due to factors like lack of communication with peers and mentors, lack of structure, or fear of judgment \cite{cardoso_inflection_2016}. These challenges manifest at both individual and group levels. In groups, complex dynamics such as hierarchical structures and the involvement of multiple stakeholders can introduce additional obstacles \cite{eden_modelling_2021}. Groupthink and fixation at the group level can stifle diverse opinions and hinder innovation, further exacerbating the difficulties faced by individual designers \cite{bang_making_2017}. These issues highlight the need for tools and methods to facilitate critical reflection and help designers overcome fixation \cite{jansson_design_1991}.

Given these challenges, various fields have explored methods to promote critical thinking. Traditional design practices and disciplines like psychology and communication studies have employed peer feedback and discussion techniques to enhance critical thinking \cite{walton_dialogue_1989}. However, these methods often rely on external facilitators or significant time investments, making them less accessible. In the HCI context, the rise of generative AI has introduced the potential of chatbots to promote critical thinking \cite{mukherjee_impactbot_2023, danry_dont_2023,liang_encouraging_2023, park_thinking_2024}. LLM-based chatbots can provide proactive, adaptive, readily available interventions, engaging users in complex, coherent conversations that challenge their assumptions and encourage deeper analysis\cite{mohammed_towards_2019, karinshak_working_2023, kasneci_chatgpt_2023, do_how_2022}.

Our work-in-progress explores generative AI, specifically large language model (LLM)-based conversational agents (CAs), to facilitate critical thinking in group design work. LLMs have demonstrated remarkable natural language understanding, generation, and dialog management capabilities, making them well-suited for facilitating reflective conversations \cite{park_thinking_2024, ma_understanding_2024}. By engaging designers in structured conversations, these agents can pose thought-provoking questions, challenge assumptions, and encourage deeper analysis of design choices. Potential use scenarios include guiding designers through self-inquiry, challenging their assumptions, and sparking new ideas to break their fixations. We will present a hypothetical mock-up and discuss potential design considerations for these systems.

This submission to the DIS workshop on "Death of the Design Researcher" aims to contribute to the discourse on the evolving role of design researchers in the age of generative AI. By exploring the potential of LLM-based CAs to facilitate critical reflection and address design fixation, we hope to suggest potential directions and design considerations for building future design tools and knowledge management methods. Our working-in-progress seeks to foster discussions on how these technologies can support the work of designers while redefining the role of design researchers in the process.

\section{Beyond Recommendations: Enhancing Critical Thinking with Generative AI}

Generative AI (GenAI)'s ability to quickly create new, realistic artifacts has opened up many opportunities for design research. Design researchers have shown creative ways to integrate GenAI into the co-creation process with designers. Image generators such as DALL-E and Midjourney have been explored for their role in supporting divergent thinking and ideas \cite{chiou_designing_2023}, as a tool to enhance architects' creativity \cite{tan_using_2024}, and for generating 2D image inspiration for 3D design \cite{liu_3dall-e_2023}. Even AI errors can inspire designers creatively \cite{liu_smart_2024}. Large language models (LLMs) can enhance the idea-generation process during the divergence stage by providing additional, diverse ideas \cite{shaer_ai-augmented_2024}. The role of LLM in a group ideation context via collaborative canvas has been discussed \cite{gonzalez_collaborative_2024}. Existing attempts to incorporate GenAI into the design process have been heavily integrated into divergent design phases, particularly those that recommend visual or conceptual references \cite{tholander_design_2023, lee_proposal_2024, shaer_ai-augmented_2024}.

However, utilizing GenAI merely as a recommender presents potential issues, such as over-reliance on AI-generated ideas \cite{bucinca_trust_2021}. This can lead to a lack of originality and depth in individual creative outputs. High exposure to AI ideas increased collective idea diversity but did not enhance individual creativity \cite{ashkinaze_how_2024}. Exposure to AI images during ideation results in more design fixation on early examples \cite{wadinambiarachchi_effects_2024}. To address these issues, AI can be leveraged to provide information or automate tasks as a collaborator that prompts deeper thought and reflection.

AI systems can enhance decision-making processes by encouraging reflective thinking; instead of providing direct answers, they can ask questions that stimulate deeper analysis \cite{danry_dont_2023}. These interactive agents have been used to provide adaptive feedback \cite{fidan_supporting_2022}) and encourage self-reflection on user performance in educational contexts \cite{mukherjee_impactbot_2023}, stimulate crowd discussion \cite{ito_agent_2022}, counter extremists \cite{blasiak_social_2021}, and question the informational validity of news content in online communities \cite{zarouali_overcoming_2021}. Because LLM-based chatbots with personas and rhetorical styles can assume multiple personas and opinions \cite{liang_encouraging_2023}, they can promote critical thinking by encouraging users to engage in discussions after consuming content, such as YouTube videos \cite{tanprasert_debate_2024}, and provide multiple perspectives to help users navigate unfamiliar decision-making scenarios \cite{park_choicemates_2023}.

Although these approaches have shown promise, it is important to note that they have primarily focused on individual contexts and have not extensively addressed group decision-making scenarios. Expanding this perspective to include group dynamics could further enhance AI systems' collaborative and reflective capabilities. Group settings introduce complexities such as hierarchy, groupthink, and peer pressure, which can affect AI interventions. They can provide more effective support by adapting AI systems to account for these social dynamics. This includes adjusting the timing and nature of AI prompts to suit group interactions and ensuring AI can dynamically respond to evolving group discourse. Developing systems that integrate seamlessly into collaborative environments can promote collective reflection and innovation, maximizing GenAI's potential in group design processes.

\section{Challenges and Opportunities of Using Conversational Agents in Group Design}






Design research and practice are collaborative and often conducted in groups rather than by individuals \cite{mathiassen_collaborative_2000}. Group dynamics must be carefully considered, involving interactions among seniors, juniors, and multiple stakeholders such as designers, developers, and planners. While group decision-making can harness collective intelligence and creativity \cite{jandric_creativity_2020, yu_collective_2012}, it also risks stifling diverse opinions due to groupthink and the spiral of silence \cite{janis_victims_1972, noelle-neumann_theory_1991}. Hierarchical structures can make it challenging for those with less authority to voice dissenting opinions, leading to potential design lock-in \cite{kennedy_hierarchical_2017}. Furthermore, groups tend to rely more heavily on AI decisions than individuals \cite{chiang_are_2023}, suggesting that group design work may use generative AI decisions as discussion starters, increasing the likelihood of design lock-in caused by generative AI in a group setting.

Conversational agents (CAs) can play various roles in group discussions, including recommender, analyst, dissenter, and facilitator \cite{ma_beyond_2024, kim_bot_2020, kim_moderator_2021, do_how_2022}. High-performing AIs are most effective as recommenders while lower-performing AIs serve better as analysts \cite{ma_beyond_2024}. Traditionally, many CAs have functioned as recommenders, providing suggestions based on data analysis and previous patterns. While effective in certain contexts, this role can inadvertently reinforce existing biases and contribute to design fixation. On the other hand, the dissenter role is particularly well-suited to promoting critical thinking by challenging prevailing opinions and encouraging diverse viewpoints. An LLM-based CA can act as a 'dissenter' in two ways: 1) providing a critical perspective on the output of other generative AIs or 2) dissenting from the group's prevailing opinion \cite{chiang_enhancing_2024}. Previous research indicates that CAs can effectively critique AI outputs but are less adept at countering prevailing group opinions \cite{chiang_enhancing_2024}. Although CAs can enhance fairness and objectivity, they often struggle with dynamic group interaction and tend to offer generalized counterarguments \cite{zheng_competent_2023}.

Implementing LLM-based agents in real-time group discussions presents challenges, such as keeping up with dynamic dialogues and responding effectively to majority opinions. Understanding the context of rapidly changing conversations and determining the appropriate timing for CA interventions are critical \cite{chiang_enhancing_2024, zheng_competent_2023}. Addressing these limitations requires further research to improve system responsiveness and adaptability. This paper introduces our LLM-based dissenter system, which is currently in the early design stages (\autoref{fig:teaser}). The system aims to build on existing research \cite{chiang_enhancing_2024} by continuously challenging dominant opinions, actively participating in real-time group discussions, and encouraging critical debate. To adapt to dynamic interactions, models may determine when to intervene in a dialogue, and techniques such as retrieval-augmented generation (RAG) can provide specific and definitive answers by consulting extensive documents or web search results \cite{khurana_why_2024}. As envisioned, this system could integrate into group discussions to offer objective, unbiased insights that promote critical and reflective thinking. While still in development, we anticipate that such a system will increase objectivity and fairness, reduce over-reliance on AI, and foster dynamic group engagement.

\begin{figure*}[]
  \centering
  \includegraphics[width=1.0\textwidth]{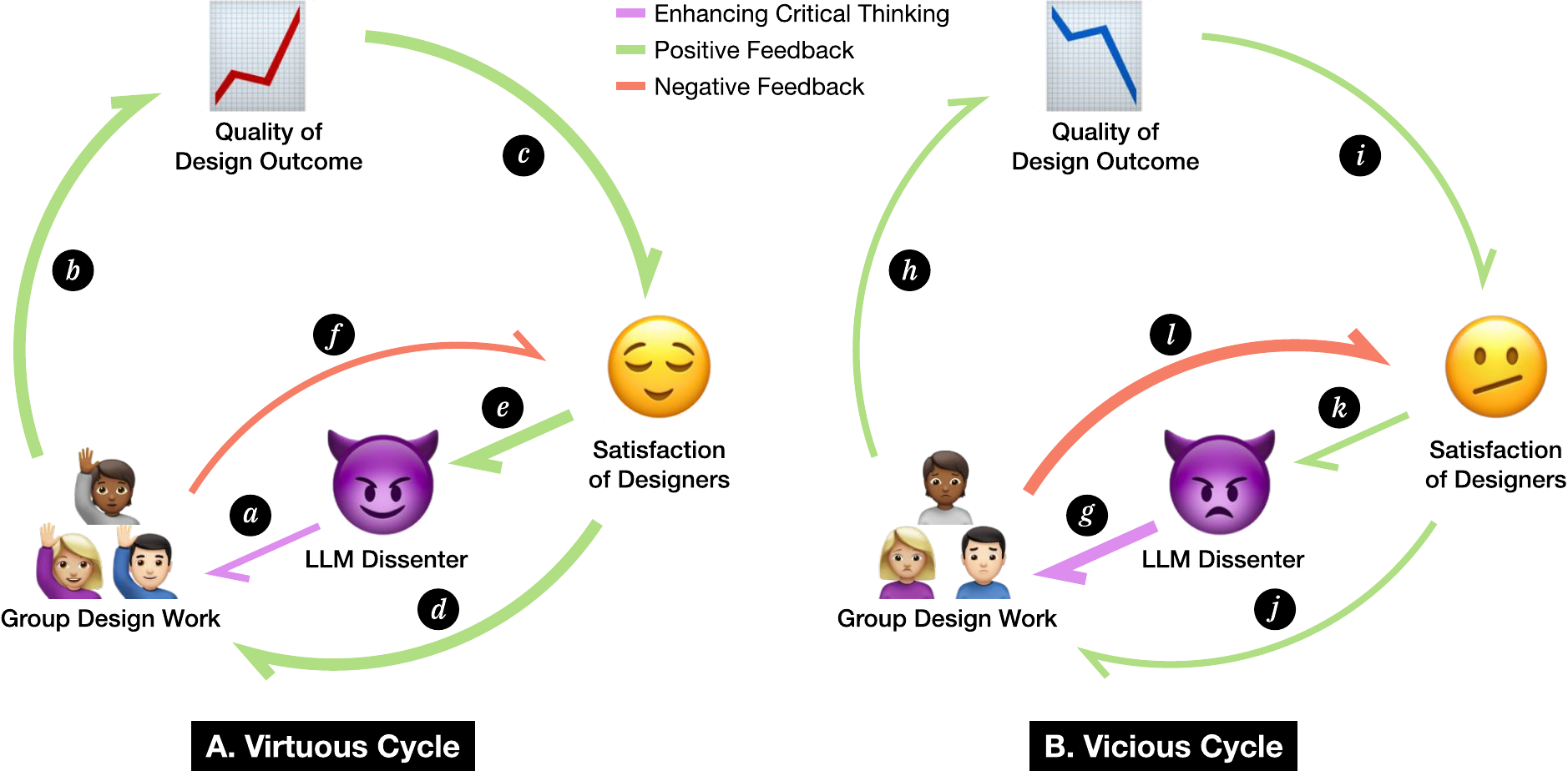}
  \caption{Hypothetical Model of the Trade-off between Critical Thinking and Designer Satisfaction in Group Design: This model illustrates how an LLM-based conversational agent (CA) acting as a naysayer can influence critical thinking and group dynamics. In a virtuous cycle, moderate stimulation of critical thinking (a) enhances design outcomes (b), increasing designers' satisfaction (c), motivating continued use of CAs (e), and fostering more critical thinking (d), with minimal negative impact (f). Conversely, in a vicious cycle, excessive stimulation (g) leads to cognitive overload and negative group dynamics (l), decreasing satisfaction (k), reducing motivation to use CAs (j), lowering design quality (h), and further diminishing satisfaction and motivation (i). This model is theoretical and has not been empirically validated.}
  \Description{}
  \label{fig:tradeOff}
\end{figure*}

\section{Potential Scenario and Applications of Conversational Agents in Group Design Process}
To illustrate the potential of LLM-based conversational agents (CAs) in facilitating critical reflection and mitigating design fixation, we present several scenarios where such systems can be effectively utilized in the design process. These scenarios highlight the role of CAs in human-AI collaboration and their ability to address issues related to design fixation and group dynamics.

\begin{itemize}
    \item \textbf{\textit{Potential Scenario 1: Enhancing Human-AI Collaboration and Mitigating Design Fixation: }} In modern design processes, generative AI is extensively used to augment human creativity by providing design recommendations and inspirations. However, this can sometimes lead to design fixation, where designers overly rely on AI-generated suggestions, limiting alternative solutions. This over-reliance can exacerbate fixation in group settings due to collective endorsement of AI suggestions. A CA, acting as a critical counterpart, can challenge these recommendations by posing alternative perspectives and encouraging designers to think beyond AI-generated ideas. This intervention reduces dependence on generative AI and mitigates group fixation caused by phenomena like groupthink and silent spirals, fostering a more diverse and innovative ideation process.
    \item \textbf{\textit{Potential Scenario 2: Facilitating Critical Reflection During Design Sprints: }} During fast-paced design sprints, teams may prioritize speed over depth, leading to superficial solutions. An LLM-based CA can integrate into the design sprint process, asking probing questions and offering critical insights at key decision points. This encourages designers to pause and reflect on their choices, enhancing the overall quality of the design outcome. The CA can serve as a continuous reflective tool, ensuring that the rapid pace does not compromise the depth and originality of the designs.
    \item \textbf{\textit{Potential Scenario 3: Supporting Novice Designers: }} Novice designers often face challenges critically evaluating their work and may overly depend on AI tools for guidance. A CA can act as a mentor, providing constructive feedback and encouraging self-reflection. By guiding novice designers through critical thinking exercises and challenging their assumptions, the CA helps build their confidence and analytical skills, making them less dependent on AI-generated recommendations and more adept at generating original ideas.
    \item \textbf{\textit{Potential Scenario 4: Real-Time Group Discussions and Decision Making: }} In real-time group discussions, keeping up with dynamic dialogues and responding effectively to majority opinions is crucial. A CA can monitor these conversations and intervene strategically to introduce alternative viewpoints and challenge dominant opinions. This dynamic interaction ensures that group decisions are well-considered and not merely a result of AI reinforcement. Techniques such as retrieval-augmented generation (RAG) can provide specific and definitive answers by consulting extensive documents or web search results, enhancing the CA's effectiveness in real-time discussions.
\end{itemize}

The primary problems we aim to address are the potential over-reliance on generative AI recommendations and the rigid group dynamics that can stifle innovation. Our work-in-progress explores the potential of LLM-based CAs to act as critical counterparts, challenging the recommendations of other generative AIs and promoting a broader exploration of design possibilities. By continuously questioning AI suggestions and fostering a culture of critical reflection, these agents can help design teams break free from fixation and enhance their creative output. Additionally, by intervening in group discussions, CAs can help counteract groupthink and encourage diverse viewpoints, addressing the issue of rigid group dynamics.

Integrating LLM-based conversational agents into the design process offers significant potential to enhance human-AI collaboration, mitigate design fixation, and address issues related to rigid group dynamics. By acting as critical counterparts and fostering deeper analysis, these agents can improve the quality and creativity of design outcomes. Our ongoing research aims to refine these systems, contributing to the development of innovative design tools that support and redefine the role of design researchers in the age of AI.

\section{Balancing Critical Thinking with Designer Satisfaction and Motivation}


Stimulating critical thinking is essential to an innovative and reflective design process. However, some trade-offs must be carefully managed to ensure a positive designer experience. While over-promoting critical thinking can lead to more diverse ideas and deeper insights in the short term, over-emphasizing it can lead to excessive cognitive load \cite{bucinca_trust_2021} and a negative impact on group dynamics in the long term \cite{chiang_enhancing_2024, do_inform_2023}. This can lower motivation and adversely affect friendly relationships among group members. Based on these existing studies, we propose a hypothetical model of the trade-off between design quality and the satisfaction and motivation of a group of designers through the promotion of critical thinking (\autoref{fig:tradeOff}).

CAs that promote critical thinking can lead to a virtuous cycle in which design quality, group satisfaction, and motivation continue to increase (\autoref{fig:tradeOff}-A). For example, if a CA in the role of a dissenter stimulates critical thinking "just right," critical thinking may be promoted as the group works on the design, leading to better design outcomes; better design outcomes may increase the satisfaction of the group of designers and motivate them to continue using the CA; and satisfaction may lead to more critical thinking in group design work. In this case, the negative effects of cognitive load or group dynamics are minimal because the CAs stimulated critical thinking 'just right,' so they don't significantly impact the satisfaction or motivation of the designer group.

On the other hand, overstimulation of critical thinking by CAs can lead to a vicious cycle of declining group satisfaction, motivation, and design quality (\autoref{fig:tradeOff}-B). If CAs 'overstimulate' critical thinking, design quality may improve in the short term. However, the cognitive load or the negative impact of group dynamics can become too much, leading to decreased satisfaction (motivation) and adversely affecting interpersonal relationships within the group. This reduces the motivation to use such CA systems and the effectiveness of group design work. As a result, the quality of the design outcome decreases, which in turn decreases the satisfaction and motivation of the designers, creating a vicious cycle.

Although this hypothetical model has yet to be validated, it is important that CAs can be utilized to promote critical thinking and group reflection without overburdening a group of designers. The focus should be encouraging designers to question assumptions and explore alternatives without putting them under undue mental strain. Finding the right balance between encouraging critical thinking and maintaining a positive designer experience is essential. Over-emphasizing or promoting critical thinking may lead to better results in the short term. Still, it can lead to a negative designer experience and a reluctance to use the system long-term. Therefore, it's important to design systems that support critical thinking while considering the appropriate designer experience based on context. By carefully designing an adaptive system that promotes critical thinking while appropriately burdening the designer, the designer group can achieve a balance that improves both the design outcome quality and the overall designer experience.
\section{Potential Design Considerations}
As we explore the potential of LLM-based conversational agents (CAs) to enhance critical reflection and mitigate design fixation in group design work, several potential design considerations could be addressed to ensure the effectiveness and acceptance of these systems.

\textbf{\textit{Timing of Interventions in Group Discussions: }}
One of the primary limitations of current CA systems is their inability to understand the real-time dynamic interactions within group discussions fully \cite{chiang_enhancing_2024, zheng_competent_2023}. This often results in delayed or poorly timed interventions, which can disrupt the flow of conversation and reduce the impact of the CA's input. Therefore, it is crucial to develop mechanisms that allow CAs to gauge the context and dynamics of group interactions accurately. This could involve real-time monitoring of conversation patterns and using advanced natural language processing techniques to determine the optimal moments for intervention. The CA should provide timely, relevant, and context-aware inputs to enhance its effectiveness in promoting critical reflection.

\textbf{\textit{Clarity and Specificity of Counterarguments: }}
Existing CAs often provide generalized responses that lack the depth needed to challenge prevailing opinions effectively. CAs should offer clearer, more detailed, pointed counterarguments to address this \cite{chiang_enhancing_2024}. This can be achieved by leveraging retrieval-augmented generation (RAG) techniques to access and present specific information from extensive databases or web resources. By providing well-substantiated and contextually relevant counterarguments, CAs can more effectively challenge assumptions and stimulate deeper critical thinking among designers.

\textbf{\textit{Adaptive Stimulation of Critical Thinking: }}
Balancing the stimulation of critical thinking with maintaining designers' satisfaction and motivation is essential. Overstimulation can lead to cognitive overload and negatively impact group dynamics, while insufficient stimulation may fail to foster meaningful reflection. An adaptive system that tailors the level of critical thinking prompts based on the context and the designers' responses can help achieve this balance. Such a system would need to continuously assess the designers' cognitive load and emotional state, adjusting its interventions to maintain optimal engagement and reflection.

\textbf{\textit{Facilitating Individual and Collective Reflection: }}
While individual reflection is crucial, collective reflection is equally important when working in teams. CAs should be capable of facilitating both types of reflection. For individual reflection, the CA can pose thought-provoking questions and provide personalized feedback. For collective reflection, it can summarize key discussion points, highlight diverse perspectives, and encourage team members to share their insights and critiques. This dual approach ensures that the benefits of reflective practice are maximized at both the individual and group levels.

\textbf{\textit{Consideration of Group Dynamics and Argumentation Styles: }}
Effective interaction within design teams requires understanding group dynamics, including the influence of ingroups and outgroups and the impact of different argumentation styles \cite{tanprasert_debate_2024}. Research has shown that these factors can significantly affect group cohesion and decision-making. CAs should be designed to adapt their argumentation styles based on the group dynamics observed. For example, a more assertive argumentation style may be effective in a highly cohesive group. In contrast, a more balanced and inclusive approach might be preferable in a diverse group with varying opinions. By adapting to the dynamic roles and styles the situation requires, CAs can better facilitate constructive and inclusive group discussions.

\textbf{\textit{Dynamic Role Adaptation: }}
CAs should not be limited to a single role, such as a dissenter, throughout the design process. Instead, they should be capable of dynamically adapting to different roles as needed, including that of a facilitator, supporter, or analyst \cite{ma_beyond_2024, kim_bot_2020, chen_integrating_2024}. This flexibility allows the CA to provide the most appropriate intervention based on the group's current needs. For instance, during the initial ideation phase, the CA might act as a facilitator to encourage various ideas. At the same time, it might adopt a more critical stance in later stages to refine and challenge the proposed solutions.

Incorporating these design considerations will enhance the effectiveness of LLM-based conversational agents in promoting critical reflection and mitigating design fixation in group design work. These systems can support innovative and reflective design processes by addressing timing, clarity, adaptability, and group dynamics. As we continue to develop and refine these CAs, it is crucial to balance stimulating critical thinking with maintaining a positive and motivating experience for designers, ensuring that these tools are both effective and sustainable in the long term.
\section{Conclusion}
This paper explores the potential of LLM-based conversational agents (CAs) to enhance human-AI collaboration, mitigate design fixation, and address rigid group dynamics in design processes. By acting as critical counterparts, these agents can challenge AI-generated recommendations, facilitate both individual and collective reflection, and dynamically adapt to various roles within design discussions. We propose key design considerations, including optimizing intervention timing, ensuring clarity in counterarguments, and balancing critical thinking with designers' satisfaction.

Our work-in-progress aims to stimulate a broader conversation about the innovative uses of generative AI in group design processes. By integrating CAs in these contexts, we hope to open new pathways for research and development, ultimately contributing to more reflective, inclusive, and effective design practices. This aligns with the workshop’s focus on the transformative potential of GenAI in reshaping design practices, enhancing creativity, and promoting ethical considerations in design research. Furthermore, it contributes to the workshop’s aim to develop practical tools and resources for design researchers, advancing the discourse on the responsible integration of AI in design research and fostering innovative, reflective design practices.



\bibliographystyle{ACM-Reference-Format}
\bibliography{DIS_wokrshop}










\end{document}